\documentclass[twocolumn,showpacs,showkeys,amsmath,amssymb,preprintnumbers,nofootinbib,floatfix,pra]{revtex4}

\usepackage{epsfig}
\usepackage{mathrsfs}
\usepackage{graphics}
\usepackage{dcolumn}
\usepackage{bm}
\usepackage{verbatim}
\usepackage[all, knot]{xy}
\xyoption{arc}
\xyoption{knot}
\xyoption{2cell}
\xyoption{poly}
\xyoption{graph}
\usepackage{color}

\begin{document}

\preprint{
%
 v1.0}

\title{Highly covariant quantum lattice gas model of the Dirac equation}

\author{Jeffrey Yepez}

\affiliation{
Air Force Research Laboratory, 29 Randolph Road, Hanscom AFB, Massachusetts 01731, USA}

\begin{abstract}
We revisit the quantum lattice gas model of a spinor quantum field theory---the smallest scale  particle dynamics 
is partitioned into unitary collide and stream operations.   The construction is covariant (on all scales down to a small length $\ell$ and small time $\tau=c\, \ell$) with respect to Lorentz transformations.  The mass $m$ and momentum $p$ of the modeled Dirac particle
depend on $\ell$ according to newfound relations $m = m_\circ\cos \frac{ 2\pi\ell}{\lambda}$ and $p = \frac{\hbar}{\ell} \sin \frac{ 2\pi\ell}{\lambda}$, respectively, where $\lambda$ is the  Compton wavelength of the modeled particle.  
These relations represent departures from a relativistically  invariant mass and the de Broglie relation---when taken as quantifying numerical errors the model is physically accurate when $\ell \ll \lambda$.   
Calculating the vacuum energy in the special case of a massless spinor field, we find that it  vanishes (or can have a small positive value) for a sufficiently large wave number cutoff.  This is a marked departure from the usual behavior of such a  massless field.
\end{abstract}



\date{November 17, 2010}

\maketitle


%
\section{Introduction}

We consider the dynamics of a spinor quantum field  where spacetime becomes discrete at scales smaller than some fundamental length.
In particular, we revisit the quantum computational lattice representation known as the quantum lattice gas model, a dynamical Feynman chessboard model of the Dirac equation \cite{feynman-cit46,feynman-65-1st-qlga}.  Variations, rediscoveries and improvements of the Feynman chessboard model have appeared over the years, 
including a model in 3+1 dimensions  by Riazanov \cite{riazanov-spj58}, an Ising spin chain representation by  Jacobson and Schulman  \cite{jacobson-jpamg.v17.84,jacobson-jpamg84},
a fundamental deterministic model by 't Hooft  \cite{JSP.53.323,FPL.10.105}, a
 lattice Boltzmann model  by Succi and Benzi quantum \cite{succi-pd93}, a unitary model by  Bialynicki-Birula \cite{bialynicki-birula-prd94},
and quantum lattice gas models  in 1+1 dimensions by Meyer \cite{meyer-JStatPhys96} and in 3+1 dimensions by this author \cite{yepez-qip-05}.
 We presently consider a  representation that retains  4-momentum conservation $E^2= (c p)^2 + (mc^2)^2$ of special relativity 
  down to a small length scale $\ell$ and time scale $\tau$.  
 The low-energy limit of the lattice model is defined as the dynamical regime where the Compton wavelength $\lambda$ of the quantum particle represented by an amplitude field $\psi(x)$ is much larger than the small scale.  $\psi(x)$ is treated as continuous  for $\lambda \ggg \ell$.  
Continuous derivatives emerge as effective quantum operators and the particle physics may be represented by the Dirac Lagrangian  ${\cal L}_\text{\tiny Dirac} = \overline{\psi} (i\gamma^\mu\partial_\mu -m_\circ)\psi$, where the Dirac matrices are $\gamma^\mu = ( \gamma^0, \gamma^i)$,  the spacetime derivatives are $\partial_\mu= ( \partial_t, \partial_i)$ for $i=1,2,3$, and $m_\circ$ is the  ``invariant" particle mass  (here expressed in natural units with $\hbar=1$ and $c=1$ for convenience).

This paper is organized as follows.    We begin in Sec.~\ref{quantum_lattice_gas_model} by formally introducing the quantum lattice gas model as a Langanian based theory. 
Then, in Sec.~\ref{proof_of_Planck_scale_covariance}, we present a mapping procedure whereby the discrete dynamics of a quantum lattice gas model is made equivalent to the Dirac equation.  This procedure leads to analytical form of the particle momentum that is a modification of the de Broglie relation. 
In Sec.~\ref{quantum_algorithm}, we present a deviation of the quantum lattice gas stream and collide operators that form the basis of our quantum algorithm for the Dirac equation.
In particular, we derive a unitary expression for the collision operator that is serves as a mechanism to give the spinor field its mass.   Our primary intent is to show that the quantum lattice gas, taken as a numerical tool for this quantum computational physics application, provides a high degree of numerical accuracy.  
Then, in Sec.~\ref{modified_de_Broglie_relation}, we examine the newfound requirements to have the dynamical equation of motion of the quantum lattice gas model  equal the Dirac equation at a selected small scale and explore the consequences of these requirements. 
We  present a calculation of the vacuum energy density of a spinor field, treating the special case of a  massless spinor field. 
Following a detailed analysis of the behavior the error terms, one finds an alternate theoretical purpose of the quantum lattice gas as a toy model.  It provides an example  where the vacuum energy of a massless spinor field can vanish or be very small.
That is, one can take the quantum lattice gas as a toy model of Planckian scale physics and thus set the small scale sizes $\ell$ and $\tau$  to the Planck length $\ell_\text{\tiny P}=\sqrt{\hbar G/c^3}$  and Planck time $\tau_\text{\tiny P}=\ell/c$,   providing a route for  a small positive cosmological constant.
In Sec.~\ref{Conclusion} is a conclusion and summary.

\section{Quantum lattice gas model}
\label{quantum_lattice_gas_model}

The proposed high-energy (small scale) quantum lattice gas representation may be formally expressed by the Lagrangian density of the form
\begin{subequations}
\label{high_energy_theory_representations}
\begin{equation}
\label{high_energy_theory}
{\cal L}^\text{\tiny grid} 
\!\!= \overline{\psi} \left[ i\gamma^0\frac{e^{\tau \partial_t} - e^{- \tau(\gamma^0\cdot\gamma^i \partial_i +im\gamma^0)}}{\tau}\right]\psi
= {\cal L}_\text{\tiny Dirac}  + {\cal O}(\tau^2).
\end{equation}
By the least action principle, this Lagrangian density leads to the equation of motion of the form
\begin{equation}
\label{high_energy_equation_of_motion}
e^{\tau \partial_t}\psi(x) = e^{- \tau\gamma^0\cdot\gamma^i \partial_i}e^{- i \tau m\gamma^0}\psi(x).
\end{equation}
Equation (\ref{high_energy_equation_of_motion}) is the equation of motion of a quantum lattice gas,  a unitary model for a system of noninteracting Dirac particles.  
On the right-hand  side of  (\ref{high_energy_equation_of_motion}), free chiral particle motion is given by a stream operator $U_\text{\tiny S}= e^{i \tau\gamma^0\cdot\gamma^i p_i}$, with  momentum operator $p_i = -i \partial_i$.  A mass-generating 
interaction that causes a lefthanded particle to flip into a right-handed particle (and vice versa) is given by a unitary collision operator $U_\text{\tiny C}=e^{-i \tau m \gamma^0 }$.
The  product $U_\text{\tiny QLG} = U_\text{\tiny S} U_\text{\tiny C}$ is the local evolution operator of a quantum lattice gas system acting on the spinor field $\psi^\text{\tiny T}(x) = 
{
 \begin{pmatrix}
      \psi_\text{\tiny L}(x)    &
       \psi_\text{\tiny R}(x)
\end{pmatrix}}$. 
The lefthand side of (\ref{high_energy_equation_of_motion}) is a newly computed state 
$\psi'(x)\equiv  e^{\tau \partial_t}\psi(x)$ at time $t+\tau$, so (\ref{high_energy_equation_of_motion}) may be written as a quantum algorithmic map
\begin{equation}
\label{high_energy_quantum_algorithm_demo}
\psi'(x)=U_\text{\tiny S} U_\text{\tiny C} \psi(x) \mapsto \psi(x),
\end{equation}
\end{subequations}
taken to be homogeneously applied at all points $\bm{x}$ of space and at all increments $t$ of time.
In natural units  ($\hbar=1$ and $c=1$), the quantum lattice gas model (\ref{high_energy_quantum_algorithm_demo}) is specified in $1+1$ dimensions by the following unitary operators:
\begin{subequations}
\label{quantum_lattice_gas_operators}
\begin{eqnarray}
U_\text{\tiny S}^z &= &e^{ i \ell p_z \sigma_z}
 \\
U_\text{\tiny C}
&=&
 \sqrt{1-m_\circ^2\tau^2} -i \sigma_x e^{i \sigma_z\,\ell p_z}\, m_\circ\tau,
\end{eqnarray}
\end{subequations}
where $m_\circ$ is the mass of the modeled Dirac particle in the low-energy limit $\ell/\lambda \sim 0$.

In the low-energy limit, the ${\cal O}(\tau^2)$  error terms on the righthand side of  (\ref{high_energy_theory}) become negligible, so $ {\cal L}^\text{\tiny grid} \sim {\cal L}_\text{\tiny Dirac}$ is covariant with respect to Lorentz transformations in this limit.
Yet, can ${\cal L}^\text{\tiny grid}$  be manifestly covariant at high-energies $ \ell/\lambda \sim 1$?
We consider how to achieve the 
 covariance of (\ref{high_energy_theory}) at a small scale: it necessarily occurs when the high-energy equation (\ref{high_energy_equation_of_motion})---or equivalently the quantum lattice gas equation (\ref{high_energy_quantum_algorithm_demo})---has the form of the Dirac equation $(\gamma^\mu p_\mu +m)\psi(x)=0$.

 The model is an expression of the simple idea of a spacetime manifold that becomes discrete below a small scale $\ell$.
The prescriptions needed to make  (\ref{high_energy_quantum_algorithm_demo}) equivalent to the Dirac equation are derived in the following section.

\section{Imposing covariance at the small scale}
\label{proof_of_Planck_scale_covariance}

Here we show that (\ref{high_energy_quantum_algorithm_demo}) in the high-energy limit can be made equivalent to the Dirac equation.
We begin with a local evolution operator as a composition of ``qubit rotations'' $U_{\hat{\bm{n}}_2}=e^{-i \frac{\beta_2}{2} \hat{\bm{n}}_2\cdot \bm{\sigma}}$ and $U_{\hat{\bm{n}}_1}=e^{-i \frac{\beta_1}{2} \hat{\bm{n}}_1\cdot \bm{\sigma}}$:
\begin{subequations}
\label{double_rot_construction}
\begin{eqnarray}
\label{double_rot_construction_a}
U_{\hat{\bm{n}}_2}(\beta_2) U_{\hat{\bm{n}}_1}(\beta_1)
\!\!
&=&
\!\!
e^{-i \frac{\beta_2}{2} \hat{\bm{n}}_2\cdot \bm{\sigma}}
e^{-i \frac{\beta_1}{2} \hat{\bm{n}}_1\cdot \bm{\sigma}}
\\
\nonumber
& = &
\cos \frac{\beta_1}{2} \cos \frac{\beta_2}{2}
- \sin \frac{\beta_1}{2} \sin \frac{\beta_2}{2} 
\hat{\bm{n}}_1 \cdot \hat{\bm{n}}_2 \\
& - &  
\nonumber
i \Big[
\sin \frac{\beta_1}{2} \cos \frac{\beta_2}{2} \hat{\bm{n}}_1
+ \cos \frac{\beta_1}{2} \sin \frac{\beta_2}{2} \hat{\bm{n}}_2
\\
&&
\label{double_rot_construction_d}
- \sin \frac{\beta_1}{2} \sin \frac{\beta_2}{2} \hat{\bm{n}}_1 \times \hat{\bm{n}}_2
\Big] \cdot \bm{\sigma},
\end{eqnarray}
\end{subequations}
where $\bm{\sigma}=(\sigma_x,\sigma_y,\sigma_z)$ is a vector of Pauli matrices, $\hat{\bm{n}}_1$ and $\hat{\bm{n}}_2$ are unit vectors specifying the respective principal axes of rotation, and $\beta_1$ and $\beta_2$ are real-valued rotation angles.\footnote{In (\ref{double_rot_construction_d}) we used the identity 
$(\bm{a} \cdot \bm{\sigma})\cdot (\bm{b} \cdot \bm{\sigma}) 
=
\bm{a} \cdot  \bm{b} + i \,\left( \bm{a} \times \bm{b} \right)\cdot \bm{\sigma}
$.}
Let us take $U_\text{\tiny S}^z=e^{-i \frac{\beta_2}{2} \hat{\bm{n}}_2\cdot \bm{\sigma}}$ as our stream operator and $U_\text{\tiny C}=e^{-i \frac{\beta_1}{2} \hat{\bm{n}}_1\cdot \bm{\sigma}}$ as our collision operator. 
 Without loss of generality, we may choose the principle axis of rotation  along the $\hat{\bm{z}}$ to generate streaming,

\begin{subequations}
\label{unitary_stream_collide_operators}
\begin{equation}
\label{unitary_stream_operator_in_p_form}
U_\text{\tiny S}^z= e^{ i \ell p_z \sigma_z/\hbar}=e^{-i\frac{\beta_2}{2}\sigma_z},
\end{equation} 
and treat the quantum algorithmic map as if it were applied in 1+1 dimensions.\footnote{
The reduction from 3+1 to 1+1 dimensions is allowed because the algorithm has the product form
\(
\label{high_energy_quantum_algorithm}
\psi'(x)= U_\text{\tiny S} U_\text{\tiny C} \psi(x) \mapsto \psi(x),
\)
where $U_\text{\tiny S} =e^{-i \frac{\pi}{4}\sigma_y}U_\text{\tiny S}^x e^{i \frac{\pi}{4}(\sigma_y+\sigma_x)}U_\text{\tiny S}^y e^{-i \frac{\pi}{4}\sigma_x}U_\text{\tiny S}^z= e^{i \tau\gamma^0\cdot\gamma^i p_i}$, with Dirac matrices $\gamma^0 = \sigma_x\otimes \bm{1}$ and $\gamma^i = i\sigma_y\otimes \sigma_i$ in the chiral representation \cite{succi-pd93,yepez-qip-05}.  Streaming in each of the spatial directions occurs independently, so for simplicity we can choose to consider a Dirac wave moving along $\hat{\bm{z}}$.
}
In this frame a general collision operator is 
\begin{equation}
\label{general_collision_operator}
U_\text{\tiny C}= e^{-i\frac{\beta_1}{2} (\alpha \sigma_x + \beta \sigma_y + \gamma \sigma_z )},
\end{equation} 
\end{subequations}
where $\alpha$, $\beta$, and $\gamma$ are real valued components subject to the constraint $\alpha^2 + \beta^2+ \gamma^2=1$.  
The unitary operators (\ref{unitary_stream_collide_operators}) are applied locally and homogeneously at all the points in the system. That is, we consider a construction whereby the two principal unit vectors specifying the axes of rotation are
\begin{eqnarray}
\label{axes_of_2_qubit_rotations}
 \hat{\bm{n}}_1 
&=& 
(\alpha, \beta, \gamma)
\qquad\qquad
 \hat{\bm{n}}_2
=
(0, 0, 1).
 \end{eqnarray}
With this choice, $\hat{\bm{n}}_1  \times  \hat{\bm{n}}_2 = (\beta, -\alpha,0)$ and $\hat{\bm{n}}_1  \cdot  \hat{\bm{n}}_2 =\gamma$, so (\ref{double_rot_construction}) is a quite general representation of a quantum lattice gas evolution operator
\begin{eqnarray}
\label{double_rot_construction_explicit}
U_\text{\tiny S}^z \, U_\text{\tiny C}
& \stackrel{(\ref{axes_of_2_qubit_rotations})}{=} &
\cos \frac{\beta_1}{2} \cos \frac{\beta_2}{2}
- \gamma\sin \frac{\beta_1}{2} \sin \frac{\beta_2}{2} 
\\
\nonumber
& - &  
i \left(
\alpha\sin \frac{\beta_1}{2} \cos \frac{\beta_2}{2}
- \beta\sin \frac{\beta_1}{2} \sin \frac{\beta_2}{2}
\right)
 \sigma_x
\\
\nonumber
& - &  
i \left(
\beta\sin \frac{\beta_1}{2} \cos \frac{\beta_2}{2}
+\alpha\sin \frac{\beta_1}{2} \sin \frac{\beta_2}{2}
\right)
 \sigma_y
\\
\nonumber
& - &  
i \left(
 \gamma\sin \frac{\beta_1}{2} \cos \frac{\beta_2}{2} 
+ \cos \frac{\beta_1}{2} \sin \frac{\beta_2}{2}
\right)
\sigma_z.
\end{eqnarray}

The Dirac equation for a relativistic quantum particle of mass $m_\circ$  may be written as
\begin{equation}
i\hbar \partial_t \psi = -c\, p_z \sigma_z  \psi + m_\circ c^2 \sigma_x \psi.
\end{equation}
Its time-difference form may be written as
\begin{equation}
\label{Dirac_equation_1_plus_1_dim_time_difference_form}
\psi' =\left(1 + \frac{i c\, p_z\tau}{\hbar} \sigma_z  - \frac{i m_\circ c^2\tau}{\hbar} \sigma_x \right)\psi,
\end{equation}
for small $\tau$ and for momentum operator $p_z = -i\hbar\partial_z$.  We may view the unitary operator acting on the right-hand side of (\ref{Dirac_equation_1_plus_1_dim_time_difference_form}) as the effective low-energy operator obtained from the quantum lattice gas operator (\ref{double_rot_construction_explicit})
\begin{equation}
\label{double_rot_construction_explicit_low_energy}
U_\text{\tiny S}^z \, U_\text{\tiny C} \xrightarrow{\text{small}~\ell} 1 + \frac{i c\, p_z\tau}{\hbar} \sigma_z  - \frac{i m_\circ c^2\tau}{\hbar} \sigma_x.
\end{equation}
To establish a correspondence between (\ref{double_rot_construction_explicit}) and (\ref{double_rot_construction_explicit_low_energy}), we simply choose the real-valued components of $\hat{\bm{n}}_1$  to satisfy the following three conditions:
\begin{subequations}
\label{n_1_unit_vector_condition}
\begin{eqnarray}
\label{n_1_unit_vector_condition_a}
 \alpha\sin \frac{\beta_1}{2} \cos \frac{\beta_2}{2} 
&-&
\beta\sin \frac{\beta_1}{2} \sin \frac{\beta_2}{2}
 =   \frac{ m_\circ c^2\tau}{\hbar}  
\\
\label{n_1_unit_vector_condition_b}
\beta\sin \frac{\beta_1}{2} \cos \frac{\beta_2}{2}
&+&
\alpha\sin \frac{\beta_1}{2} \sin \frac{\beta_2}{2}  = 
0
\\
\label{n_1_unit_vector_condition_c}
\gamma\sin \frac{\beta_1}{2} \cos \frac{\beta_2}{2} 
&+&
 \ \ 
 \cos \frac{\beta_1}{2} \sin \frac{\beta_2}{2}
 = 
-\frac{c\, p_z\tau}{\hbar} .
\end{eqnarray}
Additionally, we should respect the reality condition that $\hat{\bm{n}}_1$ have unit norm\footnote{
Alternatively, instead of (\ref{n_1_unit_vector_condition_d}), we could impose the condition that
\(
\cos \frac{\beta_1}{2} \cos \frac{\beta_2}{2}
- \gamma\sin \frac{\beta_1}{2} \sin \frac{\beta_2}{2} =1,
\)
forcing (\ref{double_rot_construction_explicit})  to be identical to  (\ref{double_rot_construction_explicit_low_energy}).  However, in this case, the resulting solution for components of $\hat{\bm{n}}_1$ has $\alpha$ imaginary, and this breaks the unitarity of $U_\text{\tiny C}$.
So, we impose (\ref{n_1_unit_vector_condition_d}) to strictly enforce unitarity.}
\begin{equation}
\label{n_1_unit_vector_condition_d}
\alpha^2 + \beta^2+ \gamma^2=1
\end{equation}
\end{subequations}
that we established above with the collision operator (\ref{general_collision_operator}).
For the sake of simplicity, let us start with a specialized construction whereby $\hat{\bm{n}}_1$ is perpendicular to $\hat{\bm{n}}_2$. 
The solution of (\ref{n_1_unit_vector_condition}) in this special case is
\begin{equation}
\label{n1_perp_to_n2_solution}
\alpha=\cos\frac{\beta_2}{2} 
\qquad
\beta = -\sin\frac{\beta_2}{2}
\qquad
\gamma=0.
\end{equation}
Inserting (\ref{n1_perp_to_n2_solution}) into (\ref{n_1_unit_vector_condition_a}) gives
\(
\label{beta_1_solution}
\sin  \frac{\beta_1}{2} =  \frac{ m_\circ c^2\tau}{\hbar},  
\)
and in turn (\ref{n_1_unit_vector_condition_c}) is
\(
\label{grid_equation_beta2_form}
 \sqrt{1-\left( \frac{ m_\circ c^2\tau}{\hbar}\right)^2} \sin \frac{\beta_2}{2}
 = 
-\frac{c\, p_z\tau}{\hbar} .
\)
In (\ref{unitary_stream_operator_in_p_form}) we  chose $ -\ell \,p_z /\hbar ={\beta_2}/{2}$, so in turn
we have
\begin{equation}
\label{grid_equation_p_form}
 \sqrt{1-\left( \frac{ m_\circ c^2\tau}{\hbar}\right)^2} \sin \frac{\ell p_z}{\hbar}
 = 
\frac{c\, p_z\tau}{\hbar} .
\end{equation}
This is a  grid equation that relates the cell sizes $\ell$ and $\tau$ to the mass and momentum of the quantum particle  in an intrinsic way.  
Equation (\ref{grid_equation_p_form}) can be interpreted as a rather fundamental relativistic relationship between particles and points.
In place of the theory of special relativity for classical particle dynamics in a continuum, here we have constructed a lattice-based version of special relativity for particle dynamics emerging at a small scale where the  spacetime foam has a regular structure.  

Let us consider some implications of (\ref{grid_equation_p_form}).
Squaring (\ref{grid_equation_p_form}) gives
\begin{subequations}
\begin{equation}
\left( \frac{\hbar}{\tau} \sin \frac{\ell p_z}{\hbar}\right)^2 - \left(m_\circ c^2  \sin \frac{\ell p_z}{\hbar}\right)^2= (c p_z)^2.
\end{equation}

Then adding $m_\circ^2c^4$ to both sides, we have
\begin{equation}
\label{mass_momentum_repartitioning}
 \left( \frac{\hbar}{\tau}\right)^2 \sin^2 \frac{\ell p_z}{\hbar}+   \left(m_\circ c^2\right)^2 \cos^2 \frac{\ell p_z}{\hbar} = 
 (cp_z)^2 + (m_\circ c^2)^2.
\end{equation}
\end{subequations}
This is a candidate grid-level relativistic energy equation that leads us to define a grid momentum $p_{z}^\text{\tiny grid}$ and a grid mass $m$ dependent on $\ell$ as follows:
\begin{eqnarray}
 p_{z}^\text{\tiny grid} & \equiv &  \frac{\hbar}{c\tau} \sin \frac{\ell p_z}{\hbar} 
\qquad
\qquad
 m 
\equiv
m_\circ \cos \frac{\ell p_z}{\hbar} 
 .
\end{eqnarray}
Hence, the lefthand side of (\ref{mass_momentum_repartitioning}) can be interpreted as a redefinition of the Dirac particle's  kinetic and rest energies. 
Inserting the de Broglie relation ($p_z=h/\lambda$ momentum eigenvalue), the grid mass and momentum become
\begin{eqnarray}
\label{grid_momentum_and_mass}
 p_{z}^\text{\tiny grid}&=& \frac{\hbar}{c\tau} \sin \frac{ 2\pi\ell}{\lambda} 
\qquad
\qquad
m 
=m_\circ \,\cos \frac{2\pi \ell}{\lambda} .
\end{eqnarray}
%

In the low-energy limit defined by $\lambda \ggg 2\pi\ell$, expanding (\ref{grid_equation_p_form}) to first order implies that the space and time cell sizes are linearly related by the speed light
\(
\ell = c \, \tau,
\)
an intuitive relationship that we expect to hold.  In the low-energy limit, (\ref{grid_momentum_and_mass}) reduces to 
\begin{eqnarray}
\label{grid_momentum_and_mass_low_energy}
  p_{z}&=& \frac{h}{\lambda} 
\qquad
\qquad
m=
 m_\circ .
\end{eqnarray}
That is, the low-energy limit of (\ref{grid_momentum_and_mass}) corresponds to a usual quantum particle with an invariant  mass that is entirely independent of the particle's momentum, and the quantum particle acts like a wave according to standard quantum mechanics.  
However, there is a marked departure from  standard quantum mechanics in the high-energy limit in the region $\lambda \lesssim 20\ell$ as shown in Fig.~\ref{MassMomentumVsWavelength}.  

\section{The algorithm in natural units}
\label{quantum_algorithm}

For expediency, let us now switch our dimensional convention to the natural units, $\hbar =1$ and $c=1$.\footnote{In the natural units $\hbar=1$ and $c=1$, length and time have like dimension of length (i.e. $[\ell]=[\tau]=L$) while mass, momentum, and energy values have like dimension of inverse length (i.e. $[m]=[p]=[E]=L^{-1}$). Any expression written in the natural units can be converted back to an expression in the dimensionful $M, L, T$ units by simply reinserting the speed of light and Planck's constant by $\ell\mapsto \frac{\ell}{c}$, $m \mapsto \frac{mc^2}{\hbar}$, $p \mapsto \frac{pc}{\hbar}$, and $E \mapsto \frac{E}{\hbar}$.}
We can write (\ref{grid_equation_p_form}) as
\begin{subequations}
\label{grid_equation_natural_unit_component_form}
\begin{eqnarray}
\label{grid_equation_natural_unit_component_form_a}
 \sqrt{1- m_\circ^2\tau^2}\, \sin(\ell p_z)
& =& 
 p_z\tau
 \\
\label{grid_equation_natural_unit_component_form_b}
 \sqrt{1- m_\circ^2\tau^2}\, \cos(\ell p_z)
 &= &
\sqrt{1- E^2\tau^2 },
\end{eqnarray}
or equivalently as
\(
\label{grid_equation_natural_unit_component_exponential_form}
e^{i \ell p_z} = \exp\left[ i \cos^{-1} \sqrt{\frac{1- E^2\tau^2 }{1- m_\circ^2\tau^2}}\right].
\)
\end{subequations}
Furthermore, our solution (\ref{n1_perp_to_n2_solution}) implies that the rotation axis for the collision operator is
\(
\hat{\bm{n}}_1= \hat{\bm{x}} \cos\ell p_z  + \hat{\bm{y}}\sin\ell p_z,
\)
and in turn this implies that the hermitian generator of (\ref{general_collision_operator})  is
\(
\hat{\bm{n}}_1\cdot\bm{\sigma}
=
\sigma_x  \cos\ell p_z  +\sigma_y\sin\ell p_z
=
\label{n_1_p_form}
\sigma_x e^{i \sigma_z\,\ell p_z}.
\)
Hence, since $\sin  \frac{\beta_1}{2} =   m_\circ \tau 
$, we can explicitly represent the collision operator  $U_\text{\tiny C}=e^{-i\frac{\beta_1}{2}\hat{\bm{n}}_1\cdot\bm{\sigma}}$ in terms of the mass and momentum of the quantum particle as
\begin{subequations}
\label{collision_operator_in_m_p_form}
\begin{eqnarray}
U_\text{\tiny C}
&=&\sqrt{1-m_\circ^2\tau^2} -i \hat{\bm{n}}_1\cdot\bm{\sigma}\, m_\circ\tau
 \\
\label{collision_operator_in_m_p_form_b}
&=&
 \sqrt{1-m_\circ^2\tau^2} -i \sigma_x e^{i \sigma_z\,\ell p_z}\, m_\circ\tau.
\end{eqnarray}
\end{subequations}
Multiplying by the stream operator $U_\text{\tiny S}^z=e^{i\ell p_z \sigma_z}$, the evolution operator (\ref{double_rot_construction_explicit}) can now be explicitly calculated
\begin{subequations}
\label{lattice_gas_operator_in_m_E_form}
\begin{eqnarray}
U_\text{\tiny S}^zU_\text{\tiny C}
\!\!\!\!\!
 & 
 =
 &
\! \!\!
 e^{i\ell p_z \sigma_z} \sqrt{1-m_\circ^2\tau^2} -i e^{i\ell p_z \sigma_z}\sigma_x e^{i \sigma_z\,\ell p_z}\, m_\circ\tau
  \qquad \ 
\\
 &=&
 e^{i\ell p_z \sigma_z} \sqrt{1-m_\circ^2\tau^2} -i \sigma_x\, m_\circ\tau
 \\
 \nonumber
 &=&
  \sqrt{1-m_\circ^2\tau^2} \cos p_z\ell+  i \sigma_z\sqrt{1-m_\circ^2\tau^2} \sin p_z\ell
  \\ 
  &&- \ i \sigma_x\, m_\circ\tau
  \\
 & \stackrel{(\ref{grid_equation_natural_unit_component_form_a})}{\stackrel{(\ref{grid_equation_natural_unit_component_form_b})}{=}}&
  \sqrt{1-E^2\tau^2} +  i \sigma_z p_z\tau
- \ i \sigma_x\, m_\circ\tau
\\
 &=&
\label{lattice_gas_operator_in_m_E_form_e}
  \sqrt{1-E^2\tau^2} +  i E\tau\left(\sigma_z \frac{p_z}{E}- \sigma_x\, \frac{m_\circ}{E}\right).
\end{eqnarray}
\end{subequations}
This result leads us to define the rotation axis
\begin{equation}
\label{n_12_m_p_E_form}
\hat{\bm{n}}_{12} \equiv   - \frac{m_\circ}{E}\, \hat{\bm{x}}+\frac{p_z}{E}\, \hat{\bm{z}}.
\end{equation}
Since $(\hat{\bm{n}}_{12}\cdot\bm{\sigma})^2=1$ (an involution), we are free to write (\ref{lattice_gas_operator_in_m_E_form_e}) in a manifestly unitary form $e^{-i \frac{\beta_{12}}{2} \hat{\bm{n}}_{12}\cdot \bm{\sigma}}$ as follows:
\begin{subequations}
\label{lattice_gas_operator_in_exp_form}
\begin{eqnarray}
U_\text{\tiny S}^zU_\text{\tiny C}
 &=&
\label{lattice_gas_operator_in_exp_form_a}
 \exp\left[
 i \cos^{-1}\!\left( \sqrt{1-E^2\tau^2}\right) \hat{\bm{n}}_{12}\cdot\bm{\sigma}
  \right]
  \qquad
\\
 &\stackrel{(\ref{n_12_m_p_E_form})}{=}&
\label{lattice_gas_operator_in_exp_form_b}
 \exp\left[
 i \,\frac{\cos^{-1} \sqrt{1-E^2\tau^2}}{E} \left(\sigma_z  p_z -  \sigma_x\, {m_\circ}\right)
  \right]
  \qquad
\\
 &\cong&
\label{lattice_gas_operator_in_exp_form_c}
 e^{
\left. -i \,\ell \middle(-\sigma_z  p_z +  \sigma_x\, {m_\circ}\right)
 },
  \qquad
\end{eqnarray}
\end{subequations}
where in the last line we made the identification
\begin{equation}
\cos( E\ell) = \sqrt{1-E^2\tau^2}, 
\end{equation}
which is exact to one part in $10^{139}$ (i.e. accurate to 4th order in $\ell$).
This is equivalent to identifying the gate angle with the ratio of the small scale length to the particle energy, $-{\beta_{12}/2}=\ell/E$.
Then, since $U_\text{\tiny S}^zU_\text{\tiny C} \equiv e^{-i h^\text{\tiny grid}  \tau}$, the  high-energy Hamiltonian  may be written as
\begin{equation}
\label{high_energy_Hamiltonian_exact}
h^\text{\tiny grid} \stackrel{(\ref{lattice_gas_operator_in_exp_form_c})}{=} \left.\frac{\ell}{\tau}\middle(-p_z \sigma_z + m_\circ  \sigma_x \right).
\end{equation}
Thus, we have successfully demonstrated that the quantum lattice gas equation (\ref{high_energy_quantum_algorithm_demo}) is equivalent to the Dirac equation as the hamiltonian generating its unitary dynamics is the Dirac hamiltonian, even at the small scale.

\section{Consequences of the modified de Broglie relation}
\label{modified_de_Broglie_relation}

\begin{figure}[!h!b!t]
\begin{center}
\xy
(0,0)*{
\includegraphics[width=3.4in]{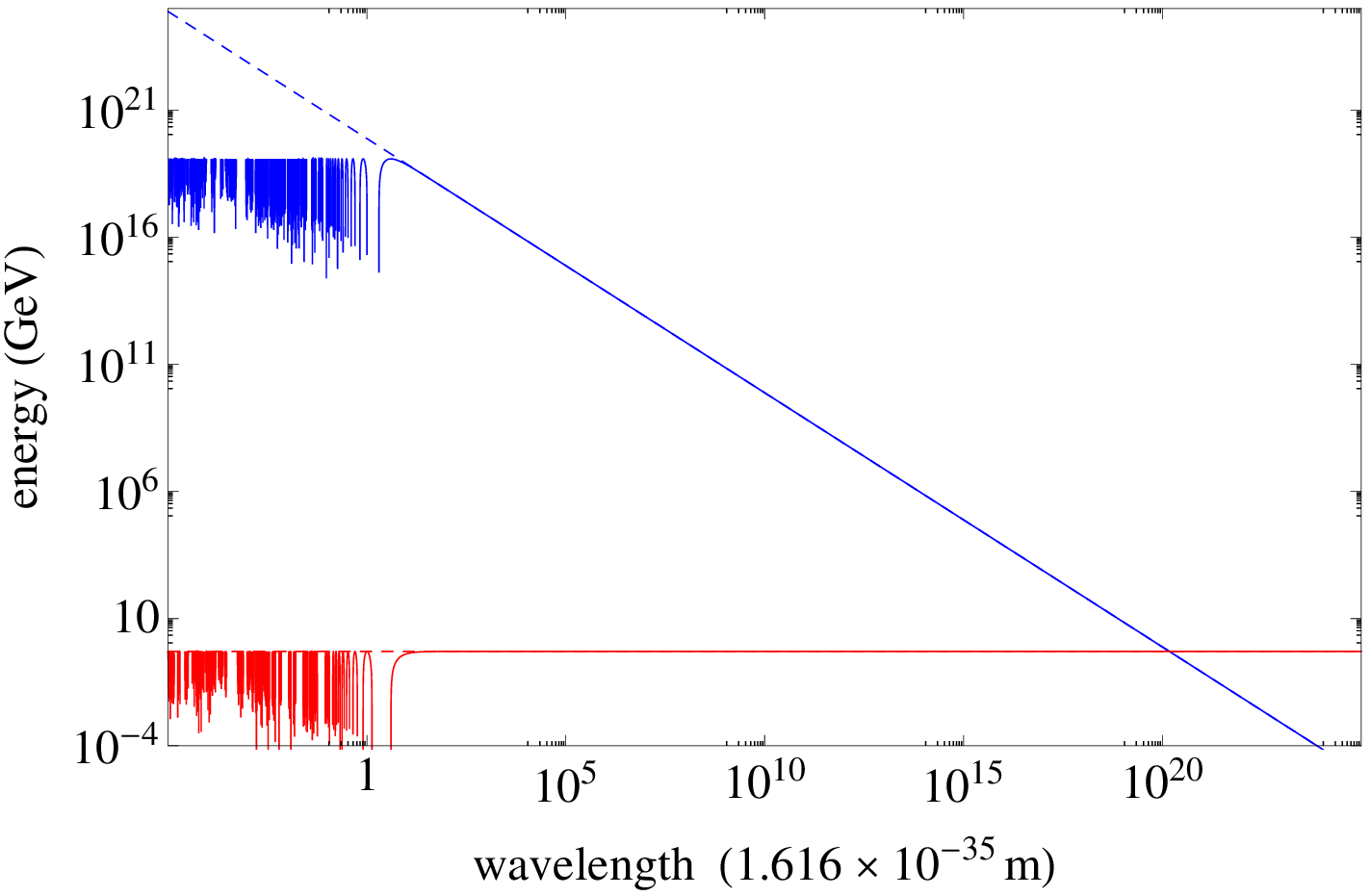}
};
(-10,-2)*{
\includegraphics[width=1.6in]{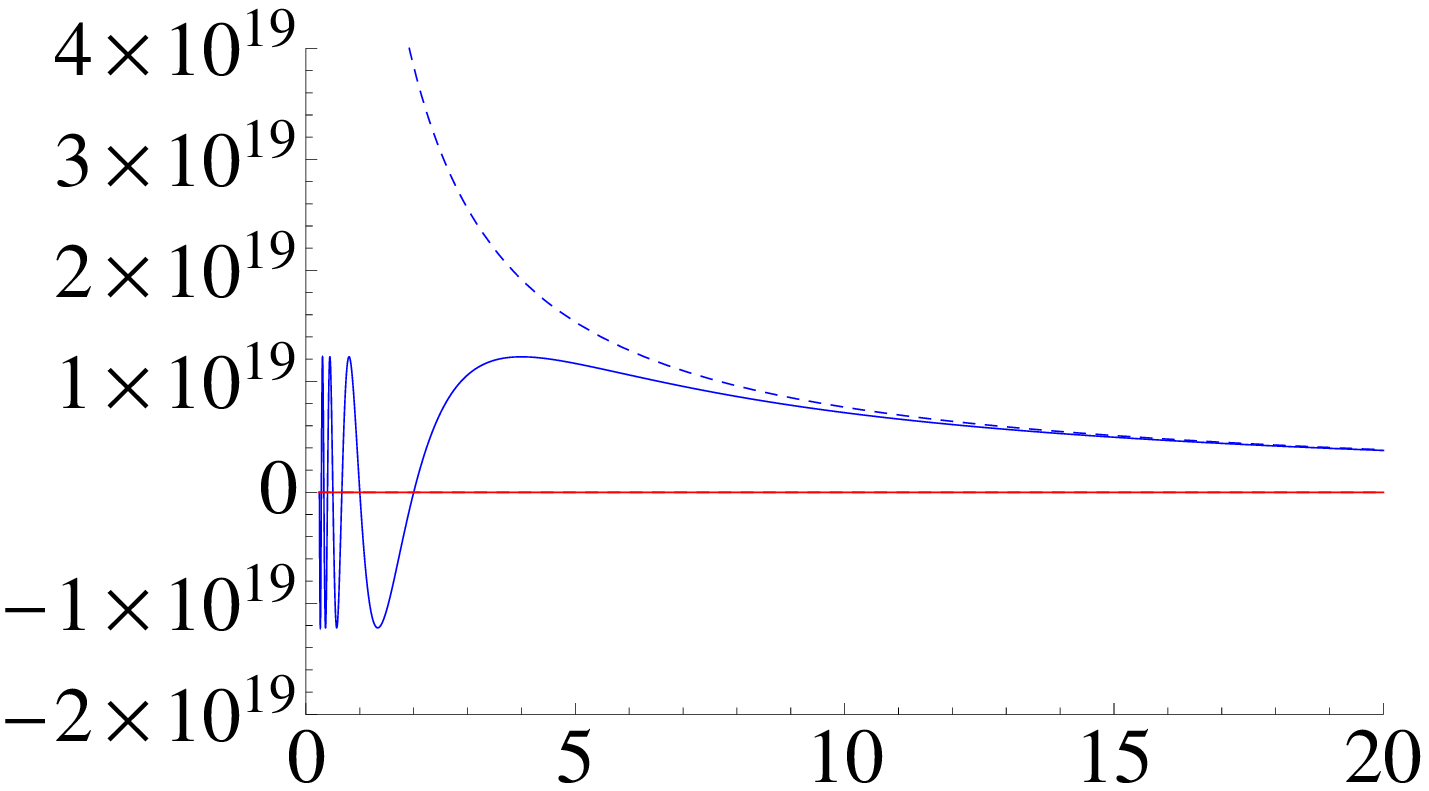}
};
(20,14)*{
\includegraphics[width=1.5in]{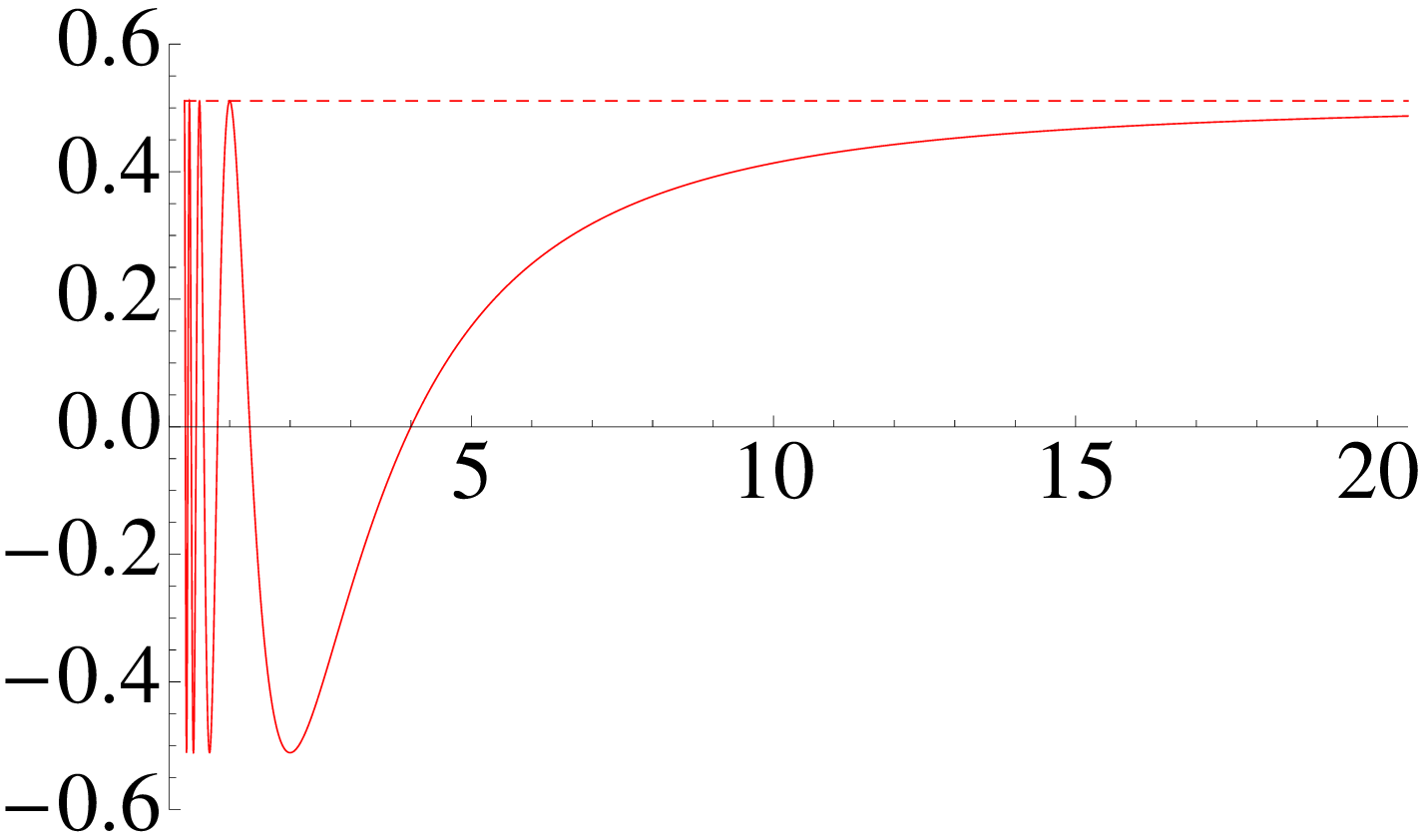}
};
\endxy
\caption{\label{MassMomentumVsWavelength} Log-log plot of (\ref{discrete_relativity_premises}) for mass (red dots) and momentum (blue dots) in GeV of a single proton versus its wavelength with the small scale set to the Planck length, $\ell \equiv \ell_\text{\tiny P}=1.616 \times 10^{-35}$m. The straight lines are the de Broglie relation  of quantum mechanics, $p=h/\lambda$ (blue dashed line), and the invariant mass of special relativity, $m_\circ=0.511$ MeV (red dashed line).  Respectively, the slopes are $-1$ and $0$ for the standard theories. 
The intersection of the mass and momentum lines occurs at 
the Compton wavelength of the Dirac particle.
The two insets are linear plots  in the extreme ultraviolet region, $\ell /4 \leq \lambda \leq 20\ell$, where the lattice theory  departs from quantum mechanics and special relativity. }
\end{center}
\end{figure}

Imposing covariance on the high-energy representation (\ref{high_energy_theory_representations}) leads to two  departures 
(\ref{grid_momentum_and_mass}),
that we derived in the previous section,  from the correct behavior given by  relativistic quantum field theory.  
First, the momentum of the quantum particle 
must obey a small length $\ell$ dependent momentum relation
\begin{subequations}
\label{discrete_relativity_premises}
\begin{equation}
\label{momentum_grid_equation}
p = \frac{\hbar}{\ell} \sin \frac{ 2\pi\ell}{\lambda} 
\end{equation}
 in place of the de Broglie relation  $p=h/\lambda$.
Second, the mass of the Dirac particle is no longer taken as an invariant quantity---it must depend on the small length as well as the particle's Compton wavelength 
\begin{equation}
\label{mass_grid_equation}
m = m_\circ\cos \frac{ 2\pi\ell}{\lambda},
\end{equation}
\end{subequations}
where $m_\circ$ is a fixed constant, otherwise interpreted in the low-energy limit as the invariant particle mass.  %
 Plots of (\ref{discrete_relativity_premises}) are given in Fig.~\ref{MassMomentumVsWavelength} for $m_\circ$ set to the electron mass.
Notice that (\ref{momentum_grid_equation}) vanishes for $\lambda=\ell$ and $\lambda=2\ell$ and oscillates about zero as $\lambda\rightarrow 0$. Also notice that (\ref{mass_grid_equation}) vanishes at $\lambda = 4 \ell$ and is in fact negative for $\lambda = 2\ell$ and  $\lambda = 3\ell$, again oscillating as $\lambda\rightarrow 0$.  
These departures from standard quantum mechanics and special relativity theory are rather consequential at very small scales, departing on scales $\lesssim 20\ell$ that can be viewed as a region where the errors in the lattice gas model are dominate.  Yet, at the  relevant larger scales (far above the small scale $\ell$), the physics of the quantum lattice gas model is indistinguishable from that predicted by the relativistic quantum field theory representation of Dirac fields.

Yet, in the context of the toy model, we should use (\ref{discrete_relativity_premises}) to calculate the vacuum energy associated with a  spin-$\frac{1}{2}$ fermion.
When we integrate over all space to determine the total density contained in a Dirac field we have
\begin{subequations}
\label{total_vacuum_density}
\begin{eqnarray}
\rho_\text{\tiny tot} 
\!\!
&=&
\!\!
 \int  \frac{d^3k}{c^2(2\pi)^3 } \sqrt{(pc)^2 + (mc^2)^2}
\\
\label{total_vacuum_density_b}
&\stackrel{(\ref{discrete_relativity_premises})}{=}&
 \frac{\hbar}{2\pi^2c\ell^4}
  \int_0^{k_c \ell} 
   d (k\ell)  ({k\ell})^2 
\sqrt{  \sin^2 k\ell + \epsilon^2 \cos^2 k\ell },
 \qquad
\end{eqnarray}
\end{subequations}
where the quantity $\epsilon \equiv{m_\circ c \ell}/{\hbar}$ is small when $m_\circ$ is much less than the  mass, $m_\circ \ll \hbar\, (\ell^2/\tau)^{-1}$.
If we take $\epsilon=0$ to model a massless relativistic particle, then
performing the integration of (\ref{total_vacuum_density_b})  yields
\begin{eqnarray}
\label{predicted_vacuum_energy}
\rho_\text{\tiny vac}^\text{\tiny theory} 
\!\!\!
&=& 
\!\!\!
  \left. \frac{\hbar}{2\pi^2c\ell^4}\left( 2 k \ell \sin k\ell -( k^2\ell^2 -2)\cos k \ell \right)\right|_0^{k_c \ell},
  \qquad
\end{eqnarray}
 which can be either positive, zero, or negative depending sensitively on the value of the wave number cutoff $k_c$ as well as on the value of the small scale $\ell$.

\begin{figure}[!t!b!h!p]
\begin{center}
\includegraphics[width=3.25in]{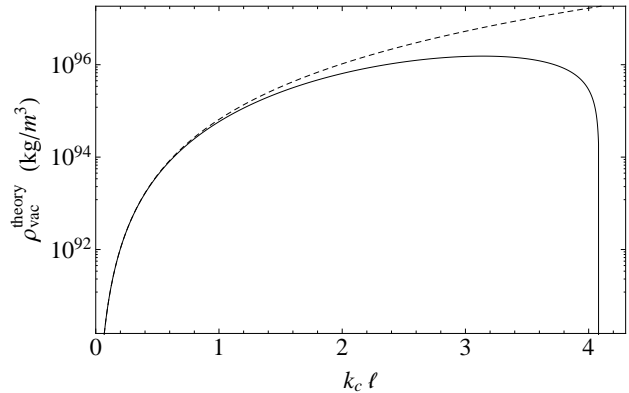}
\caption{\footnotesize Comparison of theoretical predictions of the vacuum energy density as a function of wave number cutoff times the Planck length, $k_c \ell$. The dashed curve is the standard quantum field theory prediction and the solid curve is the quantum lattice gas prediction (\ref{predicted_vacuum_energy}), where $\rho^\text{\tiny theory}_\text{\tiny vac} =0$ at  $k_c\, \ell = 4.08557$.  The theoretical prediction at $k_c \ell\sim 1$ is $10^{121}$ times too large for both curves, whereas for the solid curve $k_c \ell\gtrapprox 4$ is very close to the experimentally observed value.}
\label{VacuumEnergyPredictions}
\end{center}
\end{figure}

If one takes the quantum lattice gas as a simplistic representation of Planckian scale physics (choosing $\ell = \ell_\text{\tiny P}$) where  (\ref{discrete_relativity_premises}) is interpreted as  physical behavior instead of numerical grid error, then its prediction of an allowable small value of the vacuum energy of a massless spinor field may be viewed as a new physical mechanism.  That is, if we use (\ref{discrete_relativity_premises}) to calculate the vacuum energy associated with a  spin-$\frac{1}{2}$ fermion, then 
the toy  model can avoid the cosmological constant problem---it is not necessarily $10^{121}$ times too large.
Taking $k_c$ as a parameter in (\ref{predicted_vacuum_energy}), at  $k_c = (4.08557...)/ \ell$, we find that $\rho_\text{\tiny vac}^\text{\tiny theory}=0$, as shown in Fig.~\ref{VacuumEnergyPredictions}. 
The experimentally observed value,  $\rho_\text{\tiny vac}^\text{\tiny obs.}=9.9 \times 10^{-27} \text{kg}/\text{m}^3$ (such as obtained by  the Wilkinson Microwave Anisotropy Probe) is obtained at a slightly smaller wave number cutoff, corresponding to a grid scale about four times smaller than $\ell$,  a sub-Planckian length scale.\footnote{Our simplistic estimate is further simplified by  not considering the inflationary epoch of space under an extremely high vacuum energy density, just the dynamics of a fermionic field when the spacetime is flat with a small positive cosmological constant when its discreteness below the Planck scale becomes relevant.
}

Alternatively, one can choose the Planck scale to be smaller than the grid scale,  $\ell_\text{\tiny P} \ll \ell \ll \lambda$.  In this case, we still have $\rho_\text{\tiny vac}^\text{\tiny theory}(4.08557 k_c \ell) = 0$.  There  exists a real-valued number $C \lesssim 4.08557$ for which $\rho_\text{\tiny vac}^\text{\tiny obs.}(C k_c \ell)= \rho_\text{\tiny vac}^\text{\tiny obs.}$, although we have not predicted this number and thus do not address the fine-tunning problem.
Considerations regarding an additional fundamental length scale, in addition to the Planck scale, have recently appeared in Ref.~\cite{springerlink:10.1134/S0021364007140019}, including references therein.

\section{Conclusion}
\label{Conclusion}

We revisited the quantum lattice gas model with a unitary evolution operator $U_\text{\tiny S} U_\text{\tiny C}$ applied at a small scale $\ell$ that  advances a Dirac spinor field, represented on a grid, forward by a small time scale increment $\tau$.  We derived the conditions for which the generator of evolution is the Dirac Hamiltonian,  $U_\text{\tiny S} U_\text{\tiny C}
\cong e^{
\left. -i \,\ell \middle(-\sigma_z  p_z +  \sigma_x\, {m_\circ}\right)
 }$.
 We quantified the error of the quantum lattice gas model as a departure from standard quantum mechanical behavior for the particle momentum going as $p = (\hbar/\ell) \sin( 2\pi\ell/\lambda)$ and as its departure from a relativistically invariant particle mass going as $m = m_\circ \cos( 2\pi\ell/\lambda)$.  In this regard, the quantum lattice gas model (\ref{quantum_lattice_gas_operators}) is numerically accurate only to scales $\gtrsim 20\ell$, even though it retains covariant behavior down for scales $\gtrsim \ell$.   Yet, the numerical error of the model can be taken as a good feature---providing a mechanism for a small positive cosmological constant.

There have been a number of theoretical attempts employing, for example, supersymmetry \cite{Nilles19841}, string theory \cite{Dine1985299}, and the anthropic principle \cite{PhysRevLett.81.5501} to bridge the known chasm between the large quantum field theory prediction of $\rho_\text{\tiny vac}^\text{\tiny qft} \sim 10^{110} \text{eV}^4$ for the late-time, zero-temperature vacuum energy density of ``empty" space and the observed value of $\rho_\text{\tiny vac}^\text{\tiny obs.} \sim 10^{-11} \text{eV}^4$ associated with a small positive cosmological constant. 
The toy model presented herein gives a value of $k_c$ for which the vacuum energy vanishes. 
This is the root of the  equation $ 2 k_c \ell \sin k_c\ell -( k_c^2\ell^2 -2)\cos k_c \ell -2=0$. 
However, the theoretical considerations presented above do not tell us why the wave number cutoff should be fined tuned so that $\rho_\text{\tiny vac}^\text{\tiny theory}= \rho_\text{\tiny vac}^\text{\tiny obs.}$.\footnote{There are several routes whereby a high-energy quantum lattice gas model may predict a small positive vacuum energy density.
 At $k_c =2\pi/\ell$ we have $\rho_\text{\tiny vac}^\text{\tiny theory}  \approx - 10^{97} \text{kg}/\text{m}^3$. 
 Thus, it is possible that such a large negative $\rho_\text{\tiny vac}^\text{\tiny m=0}\lll 0$ (associated with massless chiral matter or with gauge matter comprised of paired massless fermions) can just nearly cancel a large positive $\rho_\text{\tiny vac}^\text{\tiny m>0}\ggg 0$ (associated with massive baryonic matter) leaving a small residual $\rho_\text{\tiny vac}^\text{\tiny obs}\gtrsim 0$ that is experimentally observable. At ever larger cuts, the value of the vacuum energy oscillates wildly about zero.
 }
 Nevertheless, the quantum lattice gas model appears to be one potential route to reconcile a discretized quantum field theory, at least a version modified at a small scale by (\ref{discrete_relativity_premises}), with the well accepted experimental observation of a positive cosmological constant by employing a plausible wave number cutoff parameter that corresponds to a fundamental grid scale.  In a subsequent paper, we will numerically evaluate the novel unitary collision operator (\ref{collision_operator_in_m_p_form_b}) employed in a quantum algorithmic  simulation of the dynamical behavior of a system of Dirac particles.


\section{Acknowledgements}

Thanks to G. Vahala, J. Erlich, and N.H. Margolus for helpful comments.


\end{document}